\documentstyle[11pt]{article}
\begin{document}
\title{Three--Flavor Symmetry of Hadrons Consistent With The 
Okubo-Zweig-Iizuka Rule}

\author{M. Kirchbach
\footnote{
On leave of absence from
{\it Institut f\"ur Kernphysik, Universit\"at
Mainz,  D-55099 Mainz, Germany. \/} }\\
{\it Escuela de Fisica, Univ.\ Aut.\ de Zacatecas,\/}\\
{\it Apartado Postal C-580, Zacatecas\/}\\
{\it ZAC 98068, Mexico\/}
}

\maketitle
\begin{abstract}
It is argued that the only three-flavor symmetry which  
is consistent with the Okubo-Zweig-Iizuka 
(OZI) rule, and is therefore dictated by the gluon gauge
dynamics, is the heavy
$c$ quark limit of a certain U(4)$_F$ subgroup (called S$_{OZI}$ here) 
and defined as S$_{OZI}$=
$\lim_{m_c\to \Lambda_c}$ SU(2)$_{ud}\otimes $SU(2)$_{cd}\otimes $U(1)
with the two SU(2) groups acting in turn onto the
1st and 2nd quark generations
rather than  Gell-Mann's eightfold SU(3)$_F$. Within this scheme the
presence of non-strange quarkonium components in the wave function of the
physical $\eta $ and f$_1$(1285) mesons appears necessarily  through the 
violation of the OZI rule for the pseudoscalar and axial vector mesons
by the anomalous axial U(1)$_A$ baryon number current
and not by a primordial octet flavor symmetry. 
The physical observable which selects the  
S$_{OZI}$ over the SU(3)$_F$ scheme is the non--isotriplet part
of the neutral axial current.
In the case of the S$_{OZI}$ symmetry the only non-isotriplet neutral
axial current having a well defined chiral limit is purely strange, 
while within the SU(3)$_F$ framework it is the octet axial current which
is supposed to be partially conserved. 
Accordingly, while in the latter case the
tree level $\eta N$ coupling is of significant strength, its almost
vanishes in the former, in agreement with data, 
thus confirming the relevance of the S$_{OZI}$ symmetry.

\end{abstract}

\section{Introduction}

The idea that the three-flavor symmetry of hadrons is governed
by the group SU(3)$_F$ is one of the ruling concepts of strong
interaction physics. It has its roots in the empirical observation 
that the lightest mesons constitute a family of eight
approximately mass degenerate particles of equal spins and parities
but of different electric and strangeness charges.
Same is valid for the lightest spin-1/2 baryons, while the spin-3/2
baryons join to a decuplet. To explain such patterns,
the existence of the three fundamental subbaryon degrees of freedom
with fractional electric charges, and
approximately equal masses, the $u$-, $d$-, and $s$ quarks 
(in the established notation),
has been concluded by Zweig \cite{Zweig} and Gell-Mann \cite{oway} in 
the early sixties.
Indeed, in disregrading the order of the three quarks constituting
a baryon and in using simplest combinatorics laws, one finds
ten  essentially different three quark (3q) configurations to exist.
These consist of, a) the three same--flavor combinations $uuu$, $ddd$, 
and $sss$, b) the six combinations
$uud$, $uus$, $ddu$, $dds$, $ssu$, and $ssd$ 
in which the same flavor appears only twice, and 
c) the single configuration
$uds$ where all three quarks carry different flavors.
The ten configurations are in turn identified with the spin-3/2
particles
$\Delta^{++}$, $\Delta^-$, $\Omega^-$,
$\Delta^+$, $\Sigma^*\, ^+$, $\Delta^0$, $\Sigma^*\, ^-$,
$\Xi^*\, ^0$, $\Xi^* \,^- $, and $\Sigma^*\, ^0$, respectively.  
{}From this decuplet family one easily deduces the existence of
lower dimensional pattern by first amputating the $\Delta^{++}$, 
$\Delta^-$, and $\Omega^-$ decuplet 'edges', 
then replacing $\Delta^+$ by the proton ($p$),
$\Delta^0$ by the neutron ($n$), and by removing the 
asterisk from the remaining hyperons to account for the lower spin--1/2
of the new $3q$ configurations. To obtain the octet one finally has to
account for the double occupation of the ($uds$) configuration by the
physical $\Sigma^0$ and $\Lambda $ hyperons corresponding to a neutral
charge triplet-  and a genuine charge singlet states, respectively.

The appearance of the octet and the decuplet patterns in the
spectroscopy of hadrons with low masses is therefore primarily the direct
consequence of the relevance of the three fundamental subbaryon degrees of
{}freedom. As a next step, one has to select a mathematical structure 
suited for the description of these patterns. In the early
sixties, when the knowledge about the existence of further flavors, the
quark generations, and the anomaly of the axial currents was still absent, 
it appeared natural to assume that the
three quark degrees of freedom may transform according to the fundamental
representation of the unitary
group U(3) and that the octet and the decuplet may be associated with
representations of its special form SU(3). Through this 
step, an 
ansatz for the flavor symmetry of low mass spectroscopy was made and
simultaneously  identified with the symmetry of the strong interaction 
lagrangian. 

{\it This identification is not as harmless as it looks,  
in particular because the mere existence of a regularity in the
spectroscopic patterns of  the lightest hadrons 
is not sufficient to trace back in an unique way
its symmetry origin\/}. A possible flaw of the symmetry ansatz for  
the low energy sector can easily be transferred  to both the excitation 
spectra and the dynamics of hadrons and create problems at a later stage.
Indeed, the symmetry of the lagrangian determines besides  
the properties of the states near vacuum also the hadron
excitation spectra. Moreover, under certain circumstances,
this symmetry can put strong constraints on  the hadron dynamics. 
{}For example, in the case of a symmetry realized in the Nambu--Goldstone
mode, the interaction is determined by the exchange of the corresponding 
Goldstone bosons between the matter fields with the boson-matter field
couplings being prescribed by the symmetry considered.
In assuming, therefore, the symmetry of the strong interaction lagrangian
to be SU(3)$_F$, one first expects the appearance of octet and decuplet
pattern for excited baryons. This is not unproblematic. 
It has been shown in a series of papers \cite{Ki} that the baryon
excitation spectra reveal a systematic Lorentz covariant spin clustering
with the spin clusters transforming in accordance 
to finite dimensional SU(2)$_I\otimes $O(1,3) multiplets.
This clustering of baryon resonances, in which the nucleon and 
$\Delta $ spectra below 2000 MeV are complete,
is incompatible with the 
$\lbrack $SU(3)$_F\otimes $SU(2)$_S\rbrack \times $O(3)$_L$
classification scheme with its numerous unobserved 'missing' resonances. 
A second  inevitable consequence of the 
SU(3)$_F$ symmetry ansatz concerns the chiral symmetry
SU(3)$_L\otimes $SU(3)$_R$ of QCD, the gauge theory of strong
interaction. Chiral symmetry is ordinarily supposed to be   
the global internal symmetry of the QCD lagrangian
and is considered to be realized  in the non--multiplet
Nambu--Goldstone mode with the eight lightest pseudoscalar  mesons
being associated with the expected Goldstone bosons.
Within this scheme with conserved vector currents,  the
axial ones can mostly be partially conserved if the tree level 
meson--baryon coupling constants are predetermined by means of
the corresponding Goldberger-Treiman (GT) relations \cite{ChPT}.
As long as such couplings can be extracted from either meson-nucleon
scattering cross sections, or from measuring the  meson production 
rates, off, say, a proton target, a  direct comparison to data can be 
performed. As a result of such a comparison, one finds that the 
pion--nucleon (isovector)  GT  relation is valid to quite a good  accuracy
\cite{Coon}, while the octet one, predicting the $\eta N$ coupling,
is seriously violated \cite{KiWe}.
Data are compatible with a decoupling of the $\eta $ meson from nucleons
and hyperons, rather than with its supposed octet Goldstone
boson nature (see Ref.~\cite{KiWe}) for a recent review).
The experience made with the eightfold way
over the last three decades shows that there are several aspects of hadron 
physics which appear incompatible with the  SU(3)$_F$ scheme.
Because of that, gradually and gradually the impression
arises that it is timely to revisit the SU(3)$_F$ concept by incorporating
the contemporary  knowledge about the existence of the three
quark generations, the relevance of color gauge dynamics and the 
axial current anomaly problems for hadron systems. 
In other words, it appears timely to require
three-flavor symmetry to arise from a symmetry acting within
a flavor space containing all three, or at least two, complete
quark generations after freezing out the heavy flavor degrees of freedom.
We here show that the three-flavor symmetry obtained from the four
flavor one after freezing out the $c$ quark, deviates from
the eightfold SU(3)$_F$ especially with respect to the structure
of the non--isovector part of the neutral axial current and resolves the 
problem of the observed strong suppression of the $\eta NN$ vertex over 
the $\pi NN$ one.
We find the precise three-flavor symmetry of hadrons to be
the heavy $c$ quark limit of a certain subgroup, called  S$_{OZI}$ here,
of the four-flavor unitary group U(4)$_F$, and defined as 
\begin{equation}
S_{OZI}=\lim_{m_c\to \Lambda_c} SU(2)_{ud}\otimes SU(2)_{cs}\otimes
U(1)\, ,
\label{OZI_symm}
\end{equation}
where the two SU(2) groups act in turn onto the first and second
quark generations. 

\noindent

The new  group symmetry has several advantages
over SU(3)$_F$. First of all it shares  common features with the anomaly
free SU(4) theory as it acts onto two complete quark generations.
Second, it respects the gluon gauge dynamics in so far that it
respects the Okubo-Zweig-Iizuka (OZI) rule \cite{OZI} 
by predicting the correct spectroscopic properties of the mesons in
the anomaly free vector sector.
We finally conclude that the precise
three flavor chiral symmetry of hadrons is 
S$_{OZI}^L\otimes $S$_{OZI}^R$ and  emphasize that the S$_{OZI}$  
Goldstone bosons are the pions and the kaons, while the $\eta $ meson 
behaves as a ´masked´ strange Goldstone boson, a result already 
conjectured in the previous work \cite{KiWe}.

\section{The S$_{OZI}$ group- the new three-flavor symmetry of hadrons
compatible with the OZI rule}

The main flaw  of the SU(3)$_F$ ansatz
is related to the choice for the representation
of the second diagonal SU(3)$_F$  generator, known as
the octet one, and denoted by
\begin{equation}
\lambda^8
=1/\sqrt{3} \, \mbox{diag}\,  (1,1, -2)\, .
\end{equation}
Here, $\lambda^8$ has been supposed to act onto
the three flavor space $q_3$=column (u,d,s) \cite{oway, Kaku}.
The major physical observable associated with $\lambda ^8$ 
is the wave function of the so called octet eta meson,
here denoted by $|\eta_8\rangle $, and  given below as 
\begin{equation}
|\eta_8 \rangle =  \sqrt{2}
\bar q_3 {\lambda^8 \over 2}q_3\, .  
\label{oct_eta} 
\end{equation}
The $\eta_8 $ meson is one of the three mesons occupying
the center of the pseudoscalar meson octet 
and its wave function is obtained in the course of the
orthonormalization procedure of these neutral U(3)$_F$ states.
The first such state is the neutral pion, $|\pi^0\rangle $, 
the member of the charge triplet, and  its structure
is determined in an unique way from the properties of the 
corresponding ladder operators
$T_\pm $  
\begin{equation}
\sqrt{2} |\pi^0 \rangle = T_\pm |\pi^\mp\rangle \, ,
\qquad \pi^+ =\bar d u \, , \quad \pi^-= -d\bar u\, ,
\label{ladder_pi0}
\end{equation}
as
\begin{eqnarray}
|\pi^0 \rangle =\sqrt{2} \bar q_3 {\lambda^3\over 2} q_3
&=& 
{1\over \sqrt{2}} (\bar u u -\bar d d)\, ,\nonumber\\
\lambda^3 &=& \mbox{diag} (1,-1,0)\, .
\label{pi0_wf}
\end{eqnarray}
In  presupposing the second neutral state to be the  U(1)$_F$ singlet
\begin{eqnarray}
|\eta_1^F\rangle = {1\over \sqrt{3}} 
(\bar u u + \bar d d + \bar s s)\, ,
&= &\sqrt{2}\bar q_3 {\lambda^0\over 2} q_3\, ,
\nonumber\\
\lambda^0&=& \sqrt{{2\over 3}}1\!\!1_3\, ,
\label{flv_singl}
\end{eqnarray}
the octet state in Eq.~(\ref{oct_eta}) is constructed as the only state
orthogonal to the previous two. 
In this way  Gell-Mann's eightfold way choice for the diagonal U(3)$_F$
generators $\lambda^3, \lambda^8$, and $\lambda^0$ has been made. 

This choice is by no means unique and needs be confirmed by the
data. In general, SU(N) state can equally well be described in terms of the 
new set of diagonal flavor generators 
\begin{equation}
(E_{kk})_{ij} = \delta_{ki}\delta_{kj} -
\delta_{k+1, i} \delta_{k+1, j}\, , \quad k=1,...,N-1,\quad i,j =1,...,N.
\label{Weyl_diag}
\end{equation}
For concreteness, for SU(3) one finds
E$_{11}$=diag(1,-1,0), and E$_{22}$=diag (0,1,-1). 
This is the so called Weyl's choice for the su(3)$_F$
algebra \cite{Cornw}.
Note, that the matrix $\lambda^3$ coincides with E$_{11}$ while
$\lambda^8 =$(E$_{11}$+2E$_{22})/\sqrt{3}$.
Another set of diagonal orthogonal U(3)$_F$ states can be constructed by
using instead of  the three-flavor singlet  from above,
the isospin (I) singlet
\begin{equation}
|\eta_1^I\rangle = {1\over \sqrt{2} }
(\bar u u + \bar d d )\, .
\label{eta_isosp}
\end{equation} 
Then the remaining thrid state that is orthogonal to both the
$|\pi^0\rangle $
and $|\eta_1^I\rangle $ states is the purely strange quarkonium
\begin{equation}
|\eta^s\rangle = -\bar s s\, .
\label{s_qqbar}
\end{equation}
These examples show that neither the choice for the neutral states,
nor the choice for the flavor generators is unique but it needs
further specification by comparison with data.
Mainly, one compares the masses of the mathematical with the physical
hadron states in terms of the well known ´mass formulae´ (compare
\cite{Kaku}
for details). In doing so, one finds the wave functions of the physical 
$\eta $ meson (here denoted by $|\eta \rangle $)  to slightly deviate 
from the octet state according to \cite{PART}
\begin{eqnarray}
|\eta \rangle &=& \alpha | \eta_8\rangle
-\sqrt{1-\alpha^2}|\eta_1^F\rangle \, ,\nonumber\\
 \alpha &= & \cos \, (-10.1^\circ\, )\, .
\label{lead_comp}
\end{eqnarray}
On the contrary, in the vector meson sector, no octet state is observed
at all. There, the masses of the neutral mesons  can be
explained only in
terms of a  pure separation between strange and non--strange
quarkonia according to
\begin{eqnarray}
|\phi \rangle  = -\bar s s\, ,
&\quad  &  |\omega \rangle = {1\over \sqrt{2}}
(\bar u u +\bar d d)\, , \nonumber\\
\rho^0 &=& {1\over \sqrt{2}} (\bar uu -\bar d d )\, .
\label{nonet_quark}
\end{eqnarray}

It is one of the purposes of this study to understand this contradiction.
We first show below that the flavor wave functions of the neutral vector
mesons 
are naturally obtained as U(4)$_F$ Weyl states in 
the limit of frozen charm quark degrees of freedom rather than as 
SU(3)$_F$ states. 

\noindent
Indeed, in such a case, one has to consider the three-flavor 
space to emerge from truncating the four-flavor one,
$q_4$, by treating the $c$ quark as a spectator.
The new Weyl generators in the four-flavor space following from 
Eq.~(\ref{Weyl_diag}) read \cite{KiPRD}:
\begin{eqnarray}
E_{11} =\mbox{diag} (1,-1,0,0)\, ,
&\quad & E_{22} = \mbox{diag} (0,1,-1,0)\, ,\nonumber\\ 
E_{33} = \mbox{diag} (0,0,1,-1)\, ,
&\quad & q_4 =\mbox{column} (u,d,c,s)\, .
\label{Weyls4_basis}
\end{eqnarray}
In doing so, one finds following representations of the
(orthonormalized) neutral vector meson states in the Weyl basis: 
\begin{eqnarray}
|\phi \rangle = \lim_{m_c\to \Lambda_c}\bar q_4 E_{33} q_4\, ,
&\qquad &
|\rho_0\rangle = \sqrt {2} \bar q_4 { E_{11} \over 2}q_4 \, ,
\nonumber\\
|\omega \rangle &=& \sqrt{2}\bar q_4 { {1\!\!1_2}\over 2 } q_4
\, , \quad 1\!\!1_2 = \mbox{diag} (1,1,0,0)\, .
\label{vector_sector}
\end{eqnarray}
Here the cut off parameter $\Lambda_c$ has to be sufficiently large in order
to ensure negligible $c$ quark effects on the 1 GeV scale, 
but still finite in order to preserve the anomaly free character of
the four-flavor theory \cite{ChPT,Holger}. 
Note, that $E_{11} $ is nothing but twice the third component of 
isospin, $t_3^I$, within the first quark generation,
\begin{equation}
t_3^I= {E_{11}\over 2}\, .
\label{isospin}
\end{equation} 
We further wish to introduce the quantity 
\begin{equation}
t^H_3 = {E_{33}\over 2} \, ,
\label{hyper_spin}
\end{equation}
to be called {\it hyperspin \/} in the following,
which is represented by the same matrix as weak isospin 
for the second quark generation.
As compared to the Weyl basis, the structure of the neutral vector mesons
within Gell-Mann's flavor basis appears more complicated as 
one is forced to artificially introduce the so called `magic', or,
´ideal´,  mixing angle $\theta $ according to \cite{PART}
\begin{eqnarray}
|\phi \rangle &=& \cos \theta 
\sqrt{2} \bar q_3 {\lambda^8\over 2} q_3
-\sin \theta {1\over \sqrt{3}} \bar q_3 1\!\!1_3 q_3\, ,
\nonumber\\
\theta &=& 35.3^o\, .
\label{phi_mes}
\end{eqnarray}
The last equation clearly illustrates that the 
`magic' of the ideal mixing angle is to restore Weyl's four-flavor basis
by properly back rotating the Gell-Mann's one. 
  
\noindent
On the first glance, the structure of the pseudoscalar and axial 
vector mesons seems to speak in favor of the octet way su(3)$_F$ algebra.
On the contrary, the vector mesons evidently prefer
the heavy $c$ quark limit of U(4)$_F$ in the limit of frozen charm degree
of freedom.  Simultaneously, one
immediately realizes that while in the latter case the OZI is
respected, it is strongly violated in the former. 
Indeed,  the physical  $\eta $ meson state in Eq.~(\ref{lead_comp}) 
can alternatively be reexpressed as a mixture of strange $(\bar s s)$ and 
non-strange $(\bar u u +\bar d d)$ quarkonia according to
\begin{equation}
|\eta \rangle = \cos\epsilon \, (-\bar s s) + \sin\epsilon
{1\over \sqrt{2}}(\bar u u +\bar d d)\, , \qquad
\epsilon  = 45.3^\circ\, .
\label{ss_uu}
\end{equation}
The last equation apparently violates the OZI rule
\cite{OZI} which prescribes the separation between strange and 
non--strange quarkonia. The reason is, that disconnected planar quark 
diagrams containing, say, strange to non-strange
$\bar s s \to \bar u u $  quarkonia transitions
are suppressed relative $\bar s s \to \bar s u + \bar u s $ quark
recombination diagrams. 
This follows from gluon gauge dynamics, where the  
transitions between quarkonia from the first to 
the second or third quark generations are found to
proceed over multi-gluon exchanges, 
while the recombination of a quark into two  quark-antiquark  pairs 
proceeds over single gluon exchange \cite{Badh}.
Also for the axial vector mesons f$_1$ the OZI rule
appears significantly violated. There, an angle
of $\epsilon \approx -17^\circ  $ has been reported \cite{Bolton}.
In contrast to the pseudoscalar and axial vector mesons,
Eq.~(\ref{vector_sector}) shows that 
the OZI rule is respected by the vector mesons. 
The reason for that lies in the anomaly free character of the vector meson
sector.  In contrast to that, in the pseudoscalar meson sector,
the anomalous divergency  of the axial U(1)$_A$ baryon number current
\cite{tHooft}  destroys the validity of the OZI rule, a topic 
analyzed among others in \cite{Shuryak}. 
The considerations given above lead necessarily to the insight that
Gell-Mann's three-flavor symmetry of hadrons is inconsistent with the OZI
rule and thereby with the underlying color  gauge dynamics. 
One is forced to note that, paradoxically, 
Gell-Mann's choice for the su(3)$_F$ algebra appears more
suited for describing the structure of the mesons in the anomalous axial 
current sector rather than in the anomaly free vector current one.
In connection with that, the natural question regarding the hierarchy
of the symmetries arises. 

{\it We here take the position that color
gauge symmetry has to be treated as
superior over constituent flavor symmetry and require 
the three-flavor symmetry of hadrons to 
respect the OZI rule\/}. 
Mixing up strange with non--strange quarkonia
has to emerge within this scheme through effects violating the OZI rule,
such like the effect of the 't Hooft anomaly on the structure
of the pseudoscalar and axial vector mesons.

To get an insight into the symmetry of the required properties we
consider the flavor structure of the anomaly free
electromagnetic quark current $j_\mu $ defined as 
\begin{eqnarray}
j_\mu & =&  {2\over 3}\bar u \gamma_\mu u 
               -{1\over 3} \bar d \gamma_\mu d
               +{2\over 3} \bar c \gamma_\mu c
               -{1\over 3} \bar s \gamma_\mu s\, \nonumber\\
&=&\bar q_4\,  \hat{t}^I_3 \gamma_\mu q_4 
+{1\over 2} \, \bar q_4 \hat{Y}\gamma_\mu q_4 \, ,
\quad q_4=(u\, d\, c\, s)^T\, .
\label{elm_curr}
\end{eqnarray}       
Here, we defined  $\hat{Y}$ as the genuine hypercharge operator 
\begin{eqnarray}
\hat{Y}&=&  E_{33} + \hat{B} 
 \, ,\nonumber\\
E_{33} &= & \left(\begin{array}{cccc}
0&0&0&0\\
0&0&0&0\\
0&0&\hat{C}&0\\
0&0&0&\hat{S}\end{array}\right)\, , \quad  \hat{B}=
{ {1\!\!1_{(4)} }\over 3}\, ,
\label{lambda_u4}
\end{eqnarray}
where the standard notation $\hat{B}$ has been introduced for the 
baryon number operator. 
It is remarkable that the Weyl's element E$_{33}$ 
can be identified with the difference between the hypercharge and baryon 
numbers as its action on the 
fundamental quark quadruplet $q_4$ 
reproduces charm  $(C$) and strangeness $(S$) quantum numbers of 
the quarks, once use has been made of the relations
$\hat{C}(q_4)_i=\delta_{i3}\, c$ and $\hat{S}(q_4)_j = \delta_{j4}\, s$.
{}From Eq.~(\ref{elm_curr}) one reads off that the four-flavor
electromagnetic quark current 
emerges as a Noether current with respect to the transformation
\begin{equation}
q_4\to exp\,  (\alpha \, (t^I_3+t_3^H +B/2)  )q_4\, ,
\label{group_tr}
\end{equation}
corresponding to an element of the
su(2)$_{ud}\otimes 
$su(2)$_{cs}\otimes $u(1) subalgebra of the full group U(4)$_F$
considered in the heavy $c$ quark limit.
This observation underlines once more the importance of the U(4)$_F$ 
group for hadron dynamics. Remarkably,
despite the fact, that four-flavor symmetry is more badly broken 
than the three flavor one,
as the $c$ quark is much heavier than the others, the structure of hadrons
and the transformation properties of the currents are nonetheless
determined by the representations of that very symmetry \cite{PART}.
Consequently, three-flavor currents  have to be considered 
to emerge from a flavor space containing complete
quark generations  after freezing out the heavy
quarks degrees of freedom rather than as primordial SU(3)$_F$
quantities thus requiring a revision of the traditional viewpoint
on SU(4)$_F$ as a mathematical exercise \cite{HG}.
Now, as long as $j_{\mu }$ is conserved, its total charge
\begin{equation}
Q (t) = \int j_0(t, \vec{x}\, )\mbox{d}^3\vec{x}\, ,
\label{octet_current}
\end{equation}
is a constant of motion and labels the hadron states.
When considered as an operator, $\hat{Q}$ is directly read off from
Eq.~(\ref{elm_curr})
to be related to the operators of isospin $\hat{t}^I_3 $
and hypercharge $\hat{Y}$ via the four-flavor version of the
Gell--Mann-Nakano-Nishijima relation \cite{CSY}
\begin{eqnarray}
\hat{Q} &=& \hat{t}^I_3 +{1\over 2}\hat{Y}\, \nonumber\\
&=& \hat{t}^I_3 +\hat{t}^H_3 + {1\over 2}\hat{B}\, . 
\label{GLM_NSHI}
\end{eqnarray}

Now, in the heavy $c$ quark limit of Eq.~(\ref{GLM_NSHI}),
where the charm degree of freedom is frozen out, 
one finds the expression for the hypercharge in the
truncated three-flavor space as
\begin{eqnarray}
\lim_{m_c \to \Lambda_c} 
(\hat{C} +\hat{S} + {1\over 3}\, 1\!\!1_{(4)}\,\,  ) 
=\hat{S}_{(3)} +{1\over 3}\, 1\!\!1_{(3)}\, 
&\equiv & {1\over \sqrt{3}} \lambda^8\, ,\nonumber\\
\hat{S}_{3} &=& \left(\begin{array}{ccc}
0&0&0\\
0&0&0\\
0&0&-1\end{array}\right)\, . 
\label{u4_lambda8}
\end{eqnarray}
The matrix $\hat{S}_{(3)}+ {1\over 3}\, 1\!\!1_{3}$,
on the r.h.s. of  Eq.~(\ref{u4_lambda8}),
representing the hypercharge in the 
truncated flavor space (now three dimensional) happens by accident to be
traceless despite the fact that the full four-flavor hypercharge matrix
$\hat{Y}$ in Eq.~(\ref{lambda_u4}) is not and 
coincides by chance numerically with ${1\over \sqrt{3}}\lambda^8$. 
In this way the confusing impression appears that
hypercharge can be introduced on the level of the special group SU(3)$_F$
while it is a property of the full U(4)$_F$=SU(4)$_F\otimes $U(1)
group (see \cite{Cornw}) for more details).

The physically realized  decomposition of $Q$ in Eq.~(\ref{GLM_NSHI})
suggests the precise three-flavor symmetry of hadrons to be
the heavy charm quark limit of the 
subgroup SU(2)$_{ud}\otimes$ SU(2)$_{cs}\otimes $U(1)
of U(4)$_F$ rather than Gell-Mann's SU(3)$_F$.
In the following, we will denote this group by $S_{OZI}$ and
consider it to be the precise three-flavor symmetry of hadrons, i.e.
\begin{equation}
S_{OZI} = \lim_{m_c\to \Lambda_c} 
SU(2)_{ud}\otimes SU(2)_{cs}\otimes U(1)\, .
\label{OZI_symm2}
\end{equation}
A value of $\Lambda_c \approx 1.5 $ GeV seems reasonable in that respect.
The big advantage of S$_{OZI}$ is that it allowed us to define
the quantum number of the hypercharge in 
Eq. (\ref{GLM_NSHI}) in a way consistent with the 
four-flavor Gell-Mann-Nakano-Nishijima relation, which,
in being a SU(4)$\otimes $U(1) quantity is undefinable within
the concept of a special unitary group.
Now, in contrast to $Q$, the axial flavor 
rotations corresponding to Eq.~(\ref{u4_lambda8})
\begin{equation}
q_{3}' = e^{i\beta (\hat{S}_{3} + {1\over 3}
1\!\!1_{3})\, \gamma_5}\, q_{3}\, ,
\label{no_PCAC}
\end{equation}
cannot create a conserved charge as the term containing
the unit element in the exponent in Eq.~(\ref{no_PCAC})
will give rise to the anomalously divergent axial U(1)$_A$ baryon
number current, for which no chiral limit can be formulated.
As a consequence, the full hypercharge axial current will be {\it 
anomalous\/} too. Once more, the anomaly does not disappear in going from
the four-flavor to the truncated three-flavor space. 
The only way to obtain in the chiral limit of vanishing quark 
masses a conserved, and therefore {\it observable}  
neutral flavor axial current, is to suppress the U(1)$_A$ transformation.
In doing so, one finds the following neutral axial current $j_{\mu ,5}$: 
\begin{eqnarray}
j_{\mu ,5} &=&   \bar q_4\,  \gamma_\mu\gamma_5 
\, \left( \hat{t}_3^I+ {1\over 2}(\hat{Y} -\hat{B}) \,  \right) \, q_4\, ,
\nonumber\\
&=& {1\over 2}\bar u\gamma_\mu\gamma_5 u
-{1\over 2}\bar d\gamma_\mu\gamma_5 d
+{1\over 2}\bar c\gamma_\mu\gamma_5 c
-{1\over 2}\bar s\gamma_\mu\gamma_5 s\, .
\label{weakf_curr}
\end{eqnarray}
The flavor structure of $j_{\mu , 5}$ in the last equation, which can be
partially conserved now, reflects the exclusion of the anomalous U(1)$_A$ 
current which can not be used any longer as a building block for the
construction of an anomaly free octet axial current. Remarkably,  
$j_{\mu ,5}$  respects the Okubo-Zweig-Iizuka (OZI) rule \cite{OZI}, and 
its structure is identical (up to the factor of $-1/2$) to that of the 
neutral weak axial vector current.
{}For this reason, the well established universality of the flavor changing
weak and strong axial vector currents underlying the current algebra can be 
extended to include the neutral ones.

The  $j_{\mu ,5}$  current decomposes in the four-flavor space
into an isovector ($j_{\mu , 5}^I$) 
and a purely strange SU(2)$_I$ isosinglet ($j_{\mu , 5}^{cs} $) component
while a hypercharge component is absent \cite{KiPRD}, 
\begin{eqnarray}
j_{\mu ,5} &\to  & j_{\mu ,5}^I + j^{cs} _{\mu ,5}\, , \nonumber\\
j_{\mu ,5}^I = \bar q_4 {E_{11}\over 2}\gamma_\mu \gamma_5 q_4\, , &\quad
&
j^{cs}_{\mu ,5} =
 {1\over 2} \bar q_4 \gamma_\mu\gamma_5 {E_{33}\over 2}  q_4 \, . 
\label{isov_isosc}
\end{eqnarray}
A different decomposition of $j_{\mu ,5}$ can be performed in the V-spin 
basis of the ($u,s)$ quark doublet. It is easily obtained from the last 
equation through the
replacements  $E_{11} \to E_{11} +E_{22} +E_{33}$, and $s\to d$, as 
\begin{eqnarray}
j_{\mu ,5} &\to  & j_{\mu ,5}^V + j^{cd} _{\mu ,5}\, , \nonumber\\
j_{\mu ,5}^V = \bar q_4 {{ E_{11}+E_{22}+E_{33}} \over 2}
\gamma_\mu \gamma_5 q_4\, , &\quad &
j^{cd}_{\mu ,5} = 
 \bar q_4  \gamma_\mu\gamma_5 {{-E_{22} }\over 2} q_4 \, . 
\label{Vspin_isosc}
\end{eqnarray}

Finally, in the U-spin basis of the $(d,s)$ quark doublet,
$j_{\mu ,5}$ transforms according a representation of
u(2)$_{ud}\otimes $su(2)$_{ds}$ rather than, as in the cases
before, according to the direct product of two special unitary
representations:
\begin{eqnarray}
j_{\mu ,5} &\to & 2j^I _{\mu ,5}\, +j_{\mu ,5}^U 
-j^{cu}_{\mu ,5}\, \nonumber\\
j_{\mu ,5}^U = \bar q{{ E_{22}+E_{33}} \over 2}
\gamma_\mu \gamma_5 q\, ,
&\quad &
j_{\mu ,5}^{cu} =  \bar q { {E_{11}+E_{22} } \over 2}  
\gamma_\mu\gamma_5 q\,  .
\label{Uspin_isosc}
\end{eqnarray}

Now, in the limit of frozen charm degree of freedom,
the purely charmed currents from Eqs.~(\ref{isov_isosc})-(\ref{Uspin_isosc}) 
drop out, and one finds three different 
possibilities to embed a two-flavor group within the 
restricted U(4)$_F$ group.
These are the groups 
SU(2)$_V$, SU(2)$_U$, and U(2)$_I$ with U(2)$_I$=SU(2)$_I\otimes $U(1), 
in turn corresponding to V-spin, U-spin, and isospin.
Therefore, in the case of a spectator $c$ quark, four-flavor chiral symmetry 
U(4)$^L\otimes $U(4)$^R$ restricted to
$\lbrack $SU(2)$^L_{ud}\otimes $SU(2)$^L_{cs}\otimes\-$U$^L$(1)$\rbrack 
\otimes $ 
$\lbrack $SU(2)$^R_{ud}\otimes $SU(2)$^R_{cs}\otimes$U$^R$(1)$\rbrack $ 
can be spontaneously reduced to either U(2)$_I$, 
SU(2)$_V $, or SU(2)$_U$ symmetries.
In the first case, the associated Goldstone bosons are the
three pions and the entirely non--strange isosinglet meson, 
$\eta_1^I$ from Eq.~(\ref{eta_isosp}),
while in the second and third cases one finds such to be the
two  charged ($K^+$ and $K^-$), and two neutral 
($K^0$, and $\bar K^0$) kaons, respectively. 
The broken generators $t_3^U$ and $t_3^V$ will share as a common 
neutral Goldstone boson the strange meson $|\eta ^s\rangle = -\bar s s$, 
the only neutral state that appears orthogonal to both $\pi^0$ and 
$\eta^I_1 $, respectively. In case the OZI rule was respected by the 
pseudoscalar mesons, the required strange Goldstone boson would be
$\eta^s $, the analogue to the strange $\phi $ meson 
{}from the vector meson nonet.

However, due to the influence of the U(1)$_A$ anomaly
on the spectroscopic level, the OZI rule is violated 
{}for the particular case of the pseudoscalar and axial vector mesons
and the strange and non--strange quarkonia are  mixed up \cite{Shuryak}.
This is manifest in Eq.~(\ref{lead_comp}) by means the deviation of the
wave function of the physical $\eta $ meson state from the ($-\bar s s$) 
one. Eq.~(\ref{lead_comp}) clearly illustrates that
the physical $\eta $ meson transforms according to a representation of 
$\lim_{m_c\to \Lambda_c}$ SU(2)$_{cs}\otimes $U(1)$_I$ 
rather than as a genuine SU(3)$_F$ state.
{}From this point of view, the 
significant non--strange quarkonium component of
the $\eta $ meson wave function appears as an artifact
of the violation of the OZI rule for the pseudoscalar mesons
as attributed to the U(1)$_A$ anomaly 
rather than as a consequence of an underlying fundamental SU(3)$_F$
symmetry. Nonetheless, through the significant
$\bar s s$ component in its wave function,
the $\eta $ meson is still able to meet its
purpose as the `strange' Goldstone boson mentioned above.

\section{Summary}

To summarize, we pointed out that the structure of the vector mesons
and their currents,
in respecting the fundamental gluon gauge dynamics by means of  the
OZI rule,
requires four-flavor symmetry of hadrons to be realized in the Weyl basis
and contradicts thereby Gell-Mann's eightfold  way concept.
We deduced the flavor symmetry of hadrons  on the grounds of the
transformation properties of the 
fundamental electromagnetic current in a flavor space
containing two complete quark generations with the heaviest charm flavor
being frozen out. We concluded that the only three-flavor symmetry of
hadrons of the required properties is S$_{OZI}$ in Eq.~(\ref{OZI_symm2})
rather than Gell-Mann's SU(3)$_F$.
We pointed out that the difference between the two schemes is revealed by 
the properties of the non--isovector part of the neutral 
axial current. The latter is identical to the hypercharge axial current
only for the case of the $S_ {OZI}$ symmetry and contains the 't Hooft 
anomaly. Within the S$_{OZI}$ symmetry scenario, the only isoscalar 
axial current having a well defined chiral limit is 
$\lim _{m_c\to \Lambda_c}J_{\mu ,5}^{cs}$ and purely flavored.  
On the contrary, the non--isotriplet part of the 
neutral axial current within the SU(3)$_F$ scheme is 
a F-spin scalar an if it were physically realized, it could be 
considered it as a partially conserved. 
The decision, which one of the two three-flavor
symmetries is realized in nature, can be made only through a suited 
comparison with data.    
The physical observable which is sensitive to the  
non-isotriplet neutral axial current, is the $\eta N$ coupling constant.
In the case of a (spontaneously broken) SU(3)$_\otimes $SU(3)$_R$ 
chiral symmetry, the $\eta N$ coupling  is determined  through the octet 
Goldberger-Treiman relation and appears proportional to the octet axial
coupling constant $g_A^8 = {1\over \sqrt{3}}
(\Delta u +\Delta d -2\Delta s)$ with $\Delta q_i$ standing for the
polarization of the $q_i$ quark sea.
The value of $g_A^8$ can slightly be reduced by accounting for the $\eta
-\eta '$ mixing but this reduction is still not sufficient to
explain data. Within the S$^L_{OZI}\otimes $S$^R_{OZI}$ symmetry scenario,
where a purely 
flavored isosinglet current comes in place of the octet axial current,
the (tree--level) couplings of the $\eta $ and $f_1$(1285) mesons to the 
nucleon will be proportional to 
$\lim_{\Delta c\to 0}(\Delta c -\Delta s)$, 
the negligible fraction of the 
nucleon helicity carried by the flavored  quark sea. 
This may be the main reason for the observed strong suppression
of the $\eta N$ coupling relative the value predicted by
the octet Goldberger-Treiman relation \cite{KiWe}.
To explain the small, but non--negligible 
$\eta NN$ and $f_1 NN$ vertices,  triangular
vertices of the type $a_0\pi N$ and $KK^*Y$, respectively,
have to be considered. Vertex loop corrections of that type
have been calculated, for example, in  \cite{KiTiNeu}, and shown
to be important for data interpretation.

In case of the $S_{OZI}$ symmetry and in the absence of an
anomaly, the $\eta $ boson would
appear as a purely strange quarkonium and would act as the analogue of the
$\phi $ vector meson. That this is not the case is entirely due to the 
't Hooft anomaly. Within the context of the $S_{OZI}$ symmetry,
the strong presence of non--strange quarkonium 
components in the wave function of the $\eta $ meson appeared as an
artifact of the violation of the OZI rule through the U(1)$_A$ anomaly
rather than as the consequence of an underlying fundamental 
SU(3)$_F$ symmetry.  
The consequence of the new symmetry scheme presented above is
that the spontaneous breaking of 
chiral symmetry of the QCD lagrangian will now proceed over the chain 
\begin{eqnarray}
\lim_{m_c\to \Lambda_c} ( \, 
\lbrack SU(2)^L_{ud}\otimes SU(2)^L_{cs}\rbrack
\otimes 
\lbrack SU(2)^R_{ud}&\otimes& SU(2)^R_{cs}\rbrack\,  /\nonumber\\ 
&&SU(2)_{I ,V, U } \, ,
\label{QCD_Sym}
\end{eqnarray}
rather than over $SU(3)^L\otimes SU(3)^R/SU(3)_F$.
The charged Goldstone bosons associated with 
Eq.~(\ref{QCD_Sym}) will be the three pions and the four
kaons, while an entirely strange, neutral eight Goldstone boson,
is absent but its role can still be played by
the strange quarkonium component of the $\eta $ meson.

Work supported partly by CONACyT Mexico, and partly
by the Deutsche Forschungsgemeinschaft (SFB 201).

\end{document}